\documentclass[sn-nature]{sn-jnl}

\usepackage{graphicx}
\usepackage{multirow}
\usepackage{amsmath,amssymb,amsfonts}
\usepackage{amsthm}
\usepackage{mathrsfs}
\usepackage[title]{appendix}
\usepackage{xcolor}
\usepackage{textcomp}
\usepackage{manyfoot}
\usepackage{booktabs}
\usepackage{algorithm}
\usepackage{algorithmicx}
\usepackage{algpseudocode}
\usepackage{listings}
\usepackage{geometry}
\usepackage{url}
\usepackage{pdflscape}

\renewcommand{\figurename}{Figure}

\usepackage{etoolbox}

\begin{document}

\title[Encoding gravitational waves on quantum computers]{Encoding Gravitational Waves on
Quantum Computers}

\author*[1]{\fnm{Hong} \sur{Qi}}\email{hong.qi@ligo.org}

\affil[1]{\orgdiv{School of Mathematical Sciences}, \orgname{Queen Mary University of London},\orgaddress{\street{ 327 Mile End Road}, \city{London}, \postcode{E1 4NS}, \country{United Kingdom}}}

\makeatletter
\patchcmd{\@maketitle}
  {Contributing authors:}
  {Corresponding author:}
  {}{}
\patchcmd{\@maketitle}
  {Corresponding author(s). E-mail(s):}
  {Corresponding author:}
  {}{}
\patchcmd{\authemail}{;}{}{}{}
\patchcmd{\corrauthemail}{;}{}{}{}
\makeatother

\makeatletter
\gdef\authemail{}
\gdef\corrauthemail{\textcolor{blue}{hong.qi@ligo.org}}
\makeatother

\abstract{State preparation is a bottleneck for analyzing gravitational-wave detections on quantum computers, as gravitational waves generated by colliding neutron stars comprise $\mathcal{O}(10^{5})$ to $\mathcal{O}(10^{7})$ frequency nodes. We report the first encoding of complete gravitational waveforms on quantum hardware.  Using reduced order quadrature, we can accurately represent these waveforms with only $\mathcal{O}(10^{2})$  empirical frequency nodes encoded on at most 9 qubits. Invertible classical preconditioning of waveform amplitude and phase reduces the maximum bond dimension to 2, lowering the entangling gate count for matrix product state preparation by over 30 folds. On IBM Heron processor \(\texttt{ibm\_boston}\), experiments of circuits containing 13 CNOT gates achieve mean quantum state fidelity of 0.915 for the first detected binary neutron star merger GW170817 ($\mathcal{M}_c=1.2\,M_{\odot}$) and 0.911 for a subsolar-mass binary merger ($\mathcal{M}_c=0.2\,M_{\odot}$). The corresponding waveforms comprise 257,025 and 36,782,082 frequency nodes and are represented through 135 and 478 empirical frequency nodes, achieving average waveform accuracy of 0.989 and 0.996, respectively. This method applies to arbitrary gravitational waveforms.}


\maketitle

\section{Introduction}\label{sec:intro}

Quantum algorithms for gravitational-wave (GW) detection and astrophysical inference require that gravitational waveforms are accessible as quantum states. Preparing those states is a data loading challenge, as frequency-domain waveforms from compact binary coalescences resolved in the LIGO-Virgo-KAGRA band have $N\sim 10^{3}$--$10^{8}$ frequency nodes, and $10^{5}$--$10^{8}$ for binary neutron star mergers which are among the longest signals. A vector with $N$ entries
fits on $\lceil\log_2N\rceil$ qubits, but preparing an arbitrary quantum state costs $\mathcal{O}(N)$ entangling gates \cite{Mottonen2005}, a depth over which gate faults accumulate and state fidelity falls due to hardware noise. This cost constrains the accuracies of prepared waveforms, which limits the computational capacity for downstream algorithms that use those quantum state gravitational waveforms. Exploiting structure in the waveforms, such
as sparsity or a compact tensor network representation, can reduce this preparation
cost \citep{Gonzales2025_Sparse,Berry2025_InitialState}. 

Gravitational-wave observations can be used to detect compact binary coalescences (the collisions of two compact objects like black holes and neutron stars) and, subsequently, measure their masses, spins, tidal deformabilities, internal structures, and external sky locations by comparing detector data with theoretical gravitational-wave signals predicted by General Relativity. This inference requires evaluations of millions of waveforms over a range of source parameters. Long signals are particularly demanding because they require fine frequency resolution across a broad detector band. Gravitational waveforms present a distinct
combination of two requirements: their frequency-domain amplitudes vary substantially
across the band, while their complex phases evolve through many cycles. Both
features affect the preparation circuits and the recovery of the signal from measurements. Among existing techniques, reduced order quadrature (ROQ) reliably addresses this challenge by constructing a compact yet extremely accurate representation of a waveform model
for performing the downstream astrophysical inference \citep{Field2014_Surrogate,Smith2016,Qi2021_PyROQ}.

Previous work has explored waveform quantum state encoding
in classical simulators for segments of a 2-second long analytical gravitational waveforms \citep{Hayes2023, Sarell2026}. \cite{Hayes2023} prepared a spinless
binary black hole inspiral with chirp mass $\mathcal{M}_c=28.2\,M_\odot$ on a 6-qubit
register consisting of 64 frequency points over a segment in the frequency range of $40$--$168\ \mathrm{Hz}$. Their Grover--Rudolph preparation followed
by analytic phase preparation achieved a theoretical state fidelity of
$0.995$, while their lower-gate parameterized amplitude preparation
achieved $0.979$. \cite{Sarell2026} considered only the amplitude of an analytical
inspiral waveform at 512 uniform frequency points over $40$--$170\ \mathrm{Hz}$.
Their Grover--Rudolph circuits achieved theoretical
amplitude-state fidelities of $0.99994$ and $0.99977$ at 596 and 212 CNOT gates, respectively, without encoding the waveform phase. Work on quantum hardware has focused on quantum algorithms for gravitational wave detection, with circuits using only 2 data points representing waveforms at a time for simple demonstrations  \citep{Veske2024}.   Recent LISA studies used quantum neural networks to classify 4 statistical features extracted classically,
with hardware encoding only those 4 values at a time \citep{Isfan2025_LISA, Isfan2026_Hardware}.

In this work, we encode complete, long gravitational waveforms from binary neutron star mergers on quantum hardware. We combine reduced order quadrature (ROQ) representation with classically invertible amplitude and phase
preconditioning and apply matrix product state (MPS) preparation methods. Further compilation reduces the number of physical
rotation pulses while preserving the prepared states, enabling compact MPS encoding. We study
a waveform of GW170817
\citep{Abbott2017_GW170817} and a longer subsolar waveform with chirp mass
$0.2\,M_\odot$.   Our state preparation circuits are experimented on \texttt{ibm\_boston} and \texttt{ibm\_pittsburgh} IBM quantum processors. This work establishes a procedure to prepare and read out quantum
states of encoding gravitational waveforms, laying the foundation for
quantum algorithms that perform gravitational wave detection and inference.  

\section{Results}\label{sec:results}

We first encode a waveform for GW170817, the gravitational wave from the first detected binary neutron star merger in 2017, using the waveform model IMRPhenomPv2\_NRTidalv2 \citep{Abbott2017_GW170817,Dietrich2019_NRTidalv2}. The waveform is a complex vector comprising $257{,}025$ frequency nodes (points). Direct amplitude encoding requires 18 qubits for a generic preparation circuit, and its accumulated hardware errors leave the prepared state unusable for downstream analysis. To reduce circuit complexity and enhance state fidelity and waveform accuracy, we adopt ROQ compression, amplitude and phase preconditioning, and MPS preparation. We also compare the experimental results on two IBM Heron r3 processors, namely, \texttt{ibm\_boston} and \texttt{ibm\_pittsburgh}.

\subsection{ROQ representation and exact state preparation}

We apply reduced order quadrature to the GW170817 waveform, reducing its $257{,}025$ frequency nodes to only 135 empirical frequency nodes using which the waveform can be represented accurately. Reconstructing the full waveform classically at these empirical nodes gives a waveform mismatch of
\[1-O(h_{\mathrm{ROQ}},h_{\mathrm{original}})\approx1.0\times10^{-7}\]
or smaller, where $O(h_{\mathrm{ROQ}},h_{\mathrm{original}})$ is the overlap (match) of the waveform reconstructed from ROQ bases with the original waveform. ROQ greatly reduces the dimension of the waveform vector while introducing negligible classical reconstruction error (mismatch).

We then prepare the complex vector at the 135 empirical frequency nodes in an 8-qubit register using a generic exact state preparation circuit containing 247 logical CNOT gates. On IBM Heron r3 hardware, this deep circuit produced a state fidelity of approximately 0.05 estimated from an echo and a reconstructed waveform overlap of 0.48. The waveform reconstruction uses measured node magnitudes with the known waveform phases. The loss is from errors in hardware execution and measurement.

ROQ compression reduces the dimension of the waveform vector while preserving its accuracy, but the resulting state is still costly to prepare because the 135-node complex vector retains the chirp phase winding.  Exact state preparation requires 247 logical CNOT gates. MPS approximations with maximum bond dimensions 7 and 8 achieve theoretical state fidelities of \(0.990\) and \(0.999995\), respectively, but both require 760 logical CNOT gates in our tested sequential compilation. At these circuit depths, accumulated hardware errors make the prepared states unusable for downstream analysis. We next apply amplitude- and phase-preconditioning to reduce this bond dimension and simplify the preparation circuit.

\clearpage

\begin{landscape}
\begin{figure}[p]
    \centering
    \includegraphics[
        width=\linewidth,
        height=0.99\textheight,
        keepaspectratio
    ]{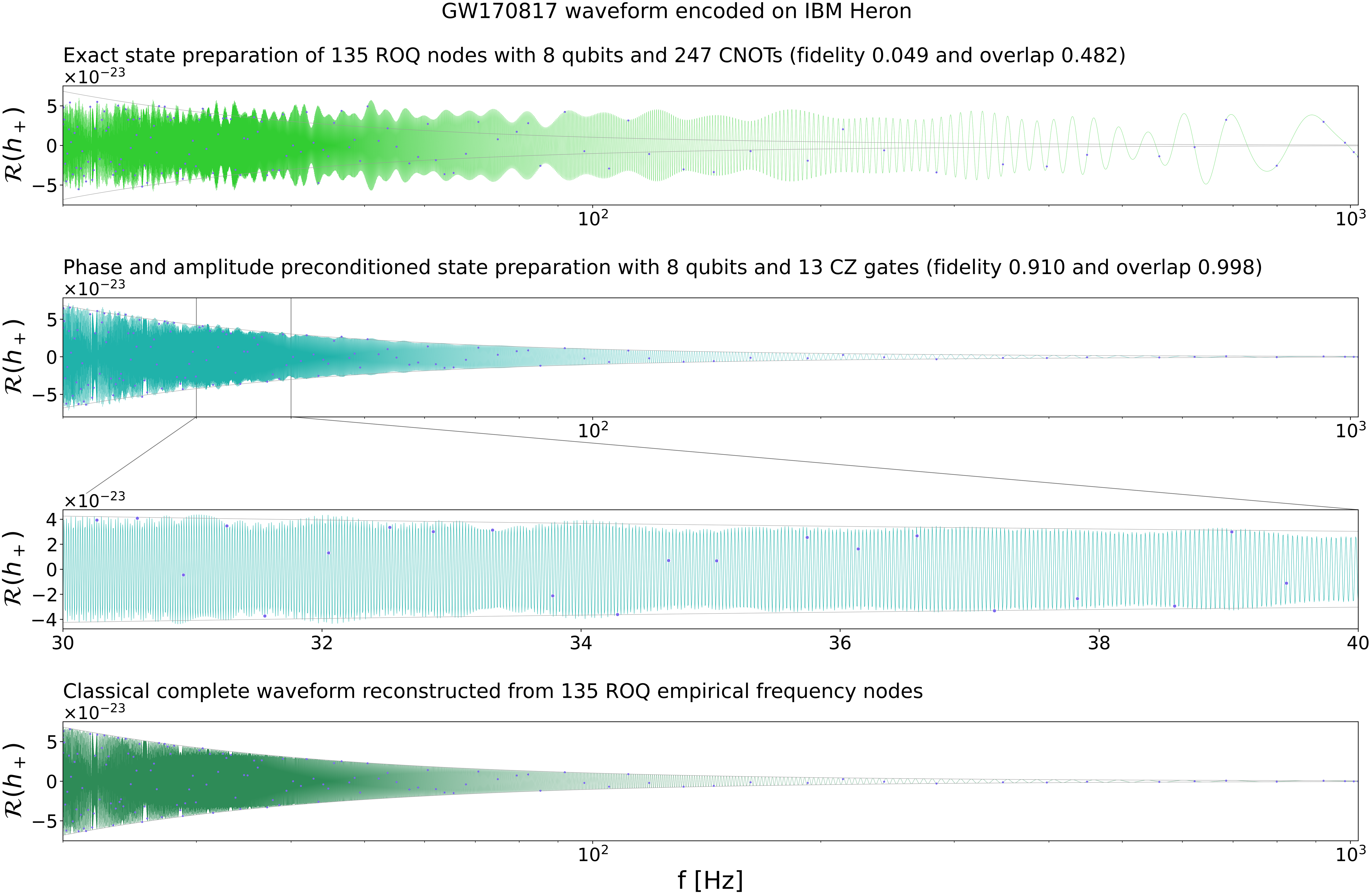}
\caption{\textbf{GW170817 waveform encoded on IBM Heron.} Results from \texttt{ibm\_boston}. From top to bottom: exact
preparation ($8$ qubits and $247$ logical CNOT gates); amplitude and
phase preconditioning with MPS preparation ($8$ qubits and $13$ physical CZ gates); a $30$--$40$~Hz zoom-in of the MPS preparation;
and the classical complete waveform from $135$ ROQ frequency nodes.
The second strip shows experiment $002$, with the highest overlap among $25$ experiments using $16{,}384$ shots per sampling and echo
circuit. Curves show the real part on the $257{,}025$-point frequency grid.
Dots are ROQ frequency nodes and gray curves show the amplitude envelope.}
\label{fig:bns}
\end{figure}
\end{landscape}

\begin{landscape}
\begin{figure}[p]
    \centering
    \includegraphics[
        width=\linewidth,    height=0.99\textheight,
        keepaspectratio
    ]{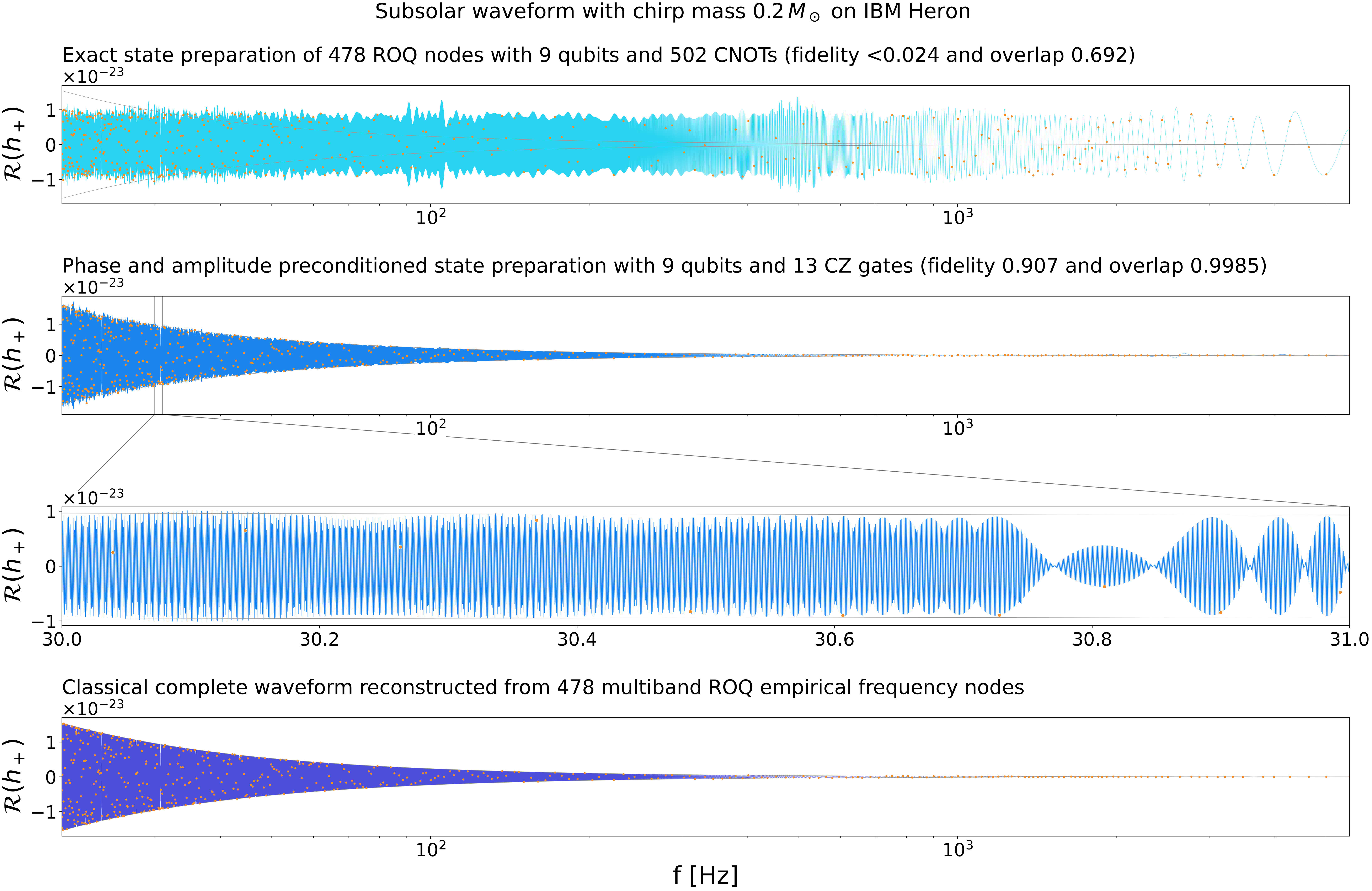}
\caption{\textbf{Subsolar waveform with chirp mass $0.2\,M_\odot$ on IBM Heron.}
State preparation results from \texttt{ibm\_boston}. From top to bottom: 1. Exact state preparation ($9$ qubits and $502$ logical CNOT gates). 2. Amplitude and phase preconditioning with MPS preparation ($9$ qubits and $13$ physical CZ gates). 3. A $30$--$31$~Hz enlargement of the second state preparation. 4.  The classical complete waveform reconstructed from $478$ multiband ROQ frequency nodes. The second strip shows experiment $005$, which has the highest waveform overlap among $25$ experiments using $65{,}536$ shots per sampling and echo circuit. Curves show the real part on the
$266{,}060$-point multiband frequencies. Dots show ROQ frequencies, and gray curves show the amplitude envelope.}
\label{fig:subsolar}
\end{figure}
\end{landscape}

\clearpage

\subsection{Preconditioning the ROQ complex vector and compact MPS preparation}

We apply both amplitude and phase preconditioning to the complex vector at the 135 ROQ empirical frequency nodes. Multiplication by $f^{7/6}$ reduces the ratio of the largest to the smallest magnitude of the 135 complex values from approximately $125$ to $1.27$, which significantly increases the measurement probabilities of the smaller magnitudes. Removing the reference chirp phase reduces the phase variation in the complex vector. In this demonstration,
the reference is the phase of the target waveform, so the transformed vector is real and non-negative. Both transformations are invertible, and their inverse factors are applied during waveform reconstruction.

\begin{figure}[t]
\centering
\includegraphics[width=\textwidth]{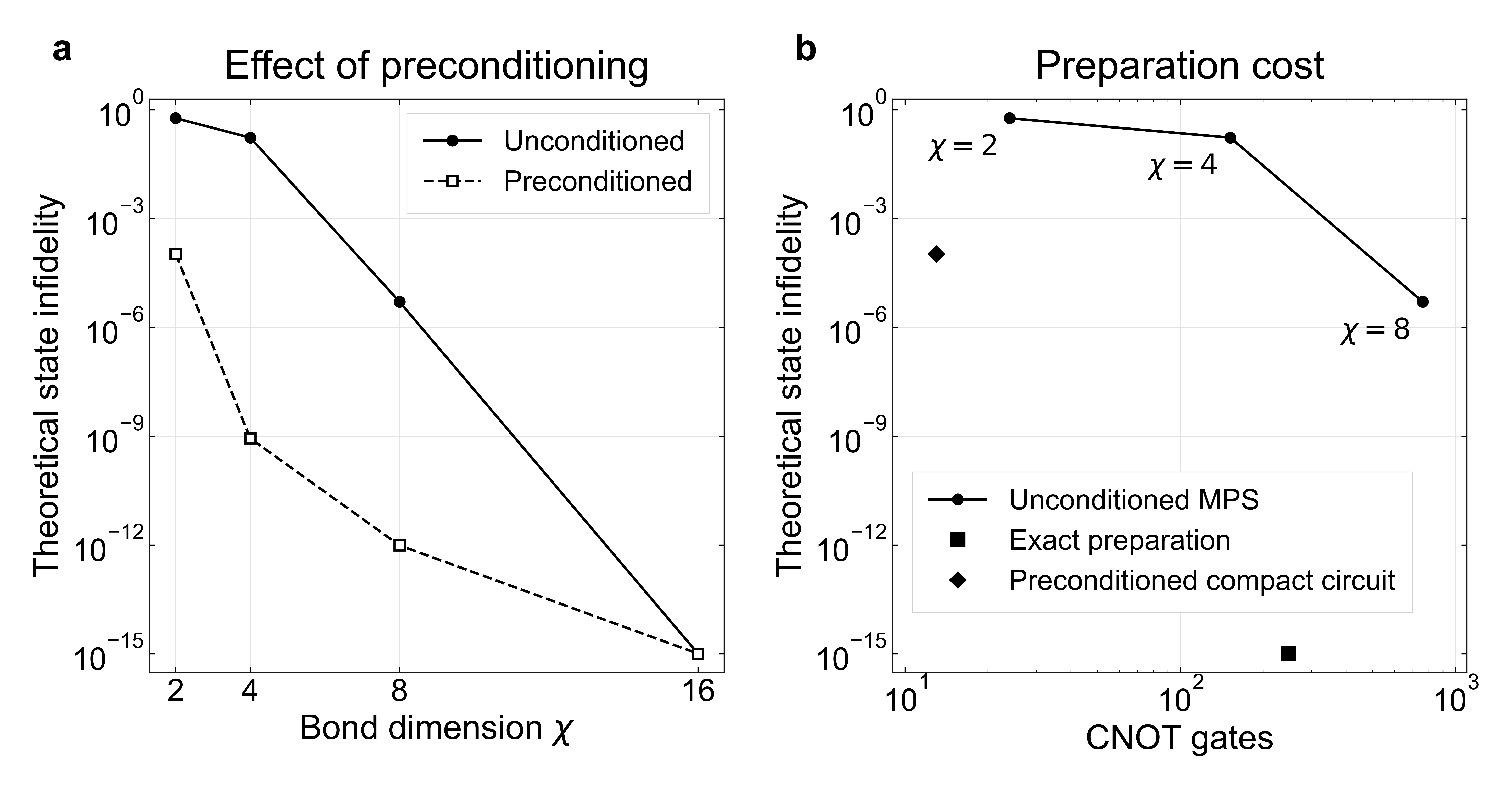}
\caption{\textbf{Effect of preconditioning on GW170817 state preparation.}
\textbf{a}, Theoretical state infidelity, $1-F$, versus maximum bond
dimension $\chi$, before and after amplitude and phase preconditioning.
\textbf{b}, Logical CNOT count and theoretical state infidelity for
unconditioned MPS preparation, exact preparation and the compact
preconditioned circuit.
Fidelities are calculated without hardware noise relative to the
corresponding unconditioned or preconditioned target vector.
Infidelities below $10^{-15}$ are displayed at $10^{-15}$.}
\label{fig:cost}
\end{figure}

The preconditioning of the complex vector enables an MPS approximation with a maximum bond dimension of 2 (reduced from 8) at a theoretical state fidelity of $0.999893$. At the same bond dimension, the unconditioned complex vector has a theoretical state fidelity of only $0.412$ (Figure~\ref{fig:cost}). The initial sequential preparation
of the preconditioned state requires 24 logical CNOT gates and 1 auxiliary bond qubit, giving 9 qubits in total. Compiling the same approximate state without a separate bond register reduces these requirements to 13 logical
CNOT gates on 8 qubits. After reversing the preconditioning and reconstructing the waveform through ROQ, the theoretical waveform accuracy is $0.999963$.

Table~\ref{tab:gw170817_preconditioning} shows the experimental performance of amplitude preconditioning, phase preconditioning, and their combination. Amplitude preconditioning alone increases the reconstructed waveform overlap from $0.482$ to $0.942$, but it retains the 247-CNOT exact preparation circuit and a low experimental state fidelity of less than $0.05$. Phase preconditioning enables the shorter 24-CNOT MPS circuit and increases the experimental state fidelity to $0.716$, but the waveform overlap is only $0.583$. For the combined preconditioning, we compile the MPS without a separate auxiliary bond register, reducing the preparation from 24 logical CNOT gates on 9 qubits to 13 logical CNOT gates on 8 qubits. Combining both preconditioning gives an experimental state fidelity of $0.718$ and a waveform overlap of $0.989$. Therefore, phase removal reduces the circuit cost, while amplitude rescaling improves the recovery of small waveform magnitudes at the ROQ frequency nodes.

We then reduce the physical rotation pulses in the hardware compilation. Relative to a direct compilation of the same logical circuit, the number of $\sqrt{X}$ pulses decreases from 63 to 37, and the number of rotation pulse layers decreases from 16 to 10. The circuit retains 13 physical CZ gates in 7 entangling layers. These changes preserve the prepared state and its theoretical fidelity to the reported precision.

Further, we simplify the 13-CNOT preparation by reducing the physical rotation pulses in its hardware compilation. The circuit still contains 13 physical CZ gates in 7 entangling layers, but compared to a direct compilation of the same logical circuit, the number of $\sqrt{X}$ pulses decreases from 63 to 37, and the number of rotation pulse layers decreases from 16 to 10. We call this compact MPS preparation. Despite these changes, both compilations with 13 CNOT gates prepare the same approximate MPS with a theoretical state fidelity of \(0.999893\).

With 25 repeated experiments on IBM \texttt{ibm\_boston} quantum processor, this circuit achieves a mean experimental state fidelity of  $0.9154\pm0.0037$, a restored ROQ node magnitude fidelity of $0.9978\pm0.0004$, and a reconstructed waveform overlap of $0.9890\pm0.0081$. The uncertainties are one standard deviations for the experiments, each using $16{,}384$ shots for the sampling circuit and the echo circuit separately. The experimental state fidelity estimated from the echo uses all echo outcomes. 
The fraction of discarded sampling outcomes decreases from $27.16\%$ for the single experiment using the sequential MPS preparation with 24 logical CNOT gates and one auxiliary bond qubit to an average of $2.00\%$ for the preparation of 25 experiments with 13 logical CNOT gates and no auxiliary qubit. Figure~\ref{fig:bns} shows the waveform reconstructed from the mean measured magnitudes at the 135 ROQ frequency nodes. 

\begin{table*}[t]
\centering
\caption{\textbf{Preparation of the GW170817 waveform at ROQ frequency nodes on \texttt{ibm\_boston}.}
CNOT counts refer to the logical preparation circuits.
$F_{\rm exp}$ denotes experimental state fidelity estimated from the echo
under the assumptions defined in Methods.
$F_{\rm node}$ is the restored magnitude fidelity at ROQ frequency nodes.  $O$ is the overlap between the waveform recovered from a hardware experiment and the ROQ representation, $O(h_{\mathrm{hardware}},h_{\mathrm{ROQ}})$. The column Discarded gives the fraction of sampling outcomes excluded
from waveform reconstruction. The echo estimate uses all echo
outcomes. Each sampling and echo circuit uses $16{,}384$ shots. The first four
rows each shows the results for one experiment, and the last row gives means and standard
deviations over 25 repeated experiments. (Prec. is short for preconditioning.)
}
\label{tab:gw170817_preconditioning}
\small
\setlength{\tabcolsep}{3pt}
\begin{tabular}{lrrrrrr}
\toprule
State Preparation
& Qubits
& CNOTs
& $F_{\rm exp}$
& $F_{\rm node}$
& $O$
& \shortstack{Discarded(\%)} \\
\midrule
exact
& 8 & 247 & 0.049 & 0.824 & 0.482 & 35.11 \\
exact with amplitude prec.
& 8 & 247 & $0.037$ & 0.986 & 0.942 & 35.94 \\
MPS with phase prec.
& 9 & 24 & 0.716 & 0.963 & 0.583 & 25.29 \\
MPS with combined prec.
& 9 & 24 & 0.718 & 0.994 & 0.989 & 27.16 \\
\midrule
compact MPS with combined 
& 8 & 13
& $0.915\pm0.004$
& $0.9978\pm0.0004$
& $0.989\pm0.008$
& $2.00\pm0.14$ \\
\bottomrule
\end{tabular}
\end{table*}

\subsection{Comparison of experiments on two quantum processors}
We compare the compact MPS preparation for GW170817 on IBM quantum processors
\texttt{ibm\_boston} and \texttt{ibm\_pittsburgh}, with 25 repeated
experiments on each processor. Both preparation circuits contain
13 physical CZ gates in 7 entangling layers and 37 $\sqrt{X}$ pulses
in 10 rotation layers. Each experiment uses $16{,}384$ shots for
the sampling circuit and the echo circuit separately.

For each processor, we select a connected chain of 8 physical qubits by minimizing the sum of calibration errors, which weigh the CZ gate errors and $\sqrt{X}$ gate errors by their occurrence counts in the preparation circuit and include readout errors once per qubit. Once selected, these 8 qubits and compiled circuits remain fixed throughout each set of 25 experiments.

\begin{table}[!ht]
\centering
\caption{
Comparison of GW170817 preparation on two IBM Heron r3 processors.
Each value includes the mean and one standard deviation with 25 experiments
per processor. Each experiment uses $16{,}384$ sampling shots and
$16{,}384$ echo shots.
}
\label{tab:gw170817_processors}
\setlength{\tabcolsep}{10pt}
\begin{tabular}{lcccc}
\toprule
Processor
& $F_{\mathrm{exp}}$
& $F_{\mathrm{node}}$
& $O$
& Discarded (\%) \\
\midrule
\texttt{ibm\_boston}
& $0.9154\pm0.0037$
& $0.9978\pm0.0004$
& $0.9890\pm0.0081$
& $2.00\pm0.14$ \\
\texttt{ibm\_pittsburgh}
& $0.9196\pm0.0068$
& $0.9978\pm0.0003$
& $0.9912\pm0.0072$
& $2.31\pm0.18$ \\
\bottomrule
\end{tabular}
\end{table}

Table~\ref{tab:gw170817_processors} compares these experimental results on both quantum processors. Overall, the two processors show similar overall performance.
The mean experimental state fidelity is slightly higher on
\texttt{ibm\_pittsburgh}, whereas \texttt{ibm\_boston} has a smaller standard deviation and a lower fraction of discarded sampling outcomes.   The fidelity of the restored waveform magnitudes at the ROQ frequency nodes averages $0.9978$ on both
processors. The mean reconstructed waveform accuracies are $0.9890$
on \texttt{ibm\_boston} and $0.9912$ on \texttt{ibm\_pittsburgh}.

At compilation, the median CZ gate error across the processor is
$0.00155$ for \texttt{ibm\_boston} and $0.00180$ for
\texttt{ibm\_pittsburgh}. During our experiments, \texttt{ibm\_boston} generally were in greater demand and had longer job queues than \texttt{ibm\_pittsburgh}, possibly for this reason.  However, this ordering is opposite to that of the measured mean state fidelities. The processor median alone
therefore does not predict the performance of these preparation
circuits. The comparison includes the selected physical qubits,
the complete compiled circuits, and their execution in separate
calibration periods.

\begin{figure}[!htbp]
\centering
\includegraphics[width=\textwidth]{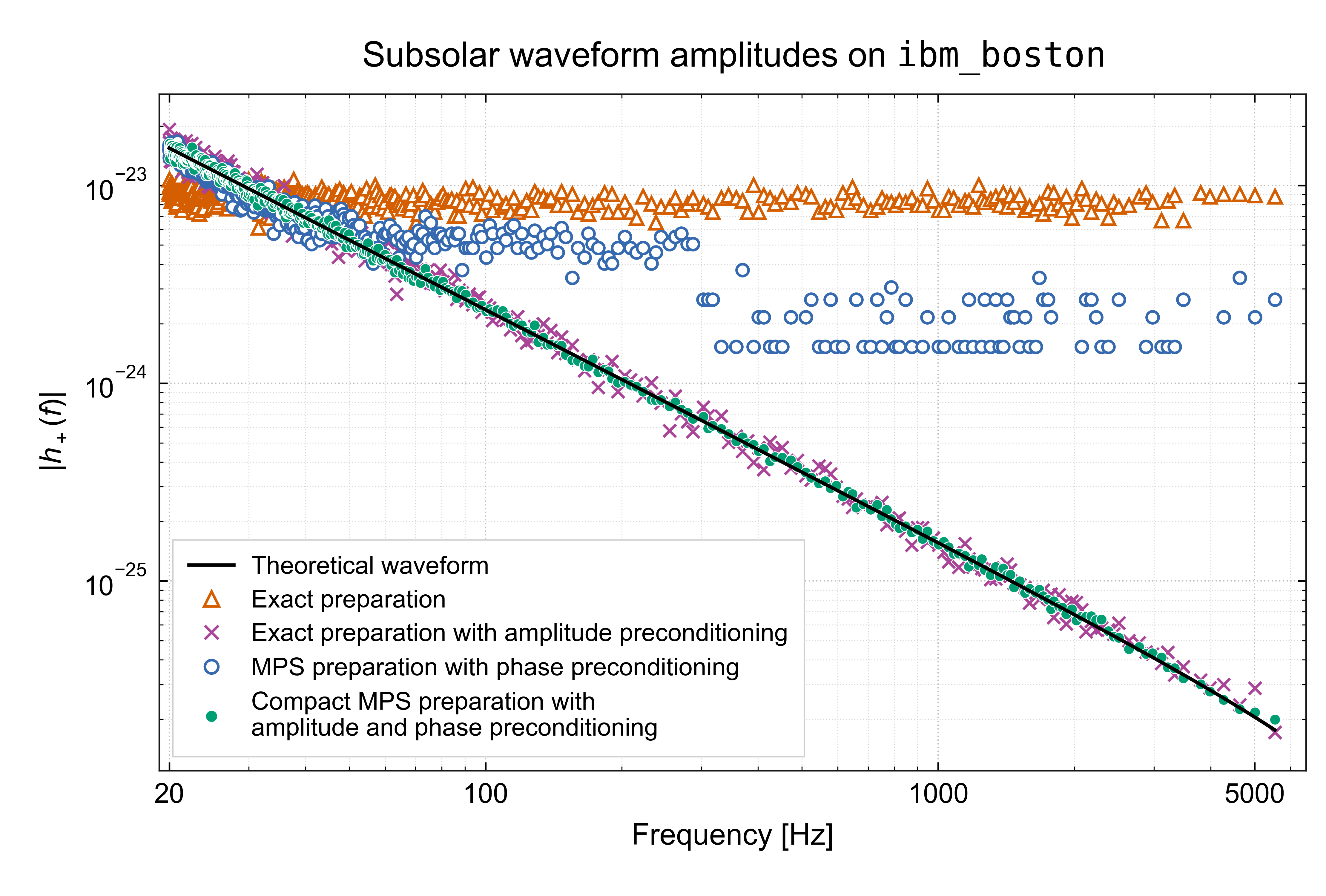}
\caption{\textbf{Subsolar waveform amplitudes on \texttt{ibm\_boston}.}
The black curve shows the theoretical waveform magnitude for
$\mathcal{M}_c=0.2\,M_\odot$. Markers show magnitudes recovered at
the 478 ROQ frequency nodes, with amplitude rescaling reversed
where applicable. Exact preparation and the individual preconditioning
methods use 16,384 shots. The compact MPS result is Experiment~008,
which has the median waveform overlap ($O=0.9973$) among
25 experiments using 65,536 shots. 
}
\label{fig:subsolar_amplitudes}
\end{figure}
We perform 25 repeated experiments on each processor at each of two
shot counts, $16{,}384$ and $65{,}536$ per sampling circuit and per
echo circuit. The physical qubits and compiled circuits remain fixed within each set of 25 experiments. 
The Subsolar sector in  Table~\ref{tab:subsolar_results}
reports the statistical results of these experiments, including the means and standard deviations.
On both processors, the experiments with $65{,}536$ shots achieve
higher mean reconstructed waveform accuracies and smaller standard
deviations than those with $16{,}384$ shots. The mean waveform overlap increases
from $0.988$ to $0.996$ on \texttt{ibm\_boston} and from $0.986$
to $0.997$ on \texttt{ibm\_pittsburgh}. At the higher shot count,
\texttt{ibm\_boston}  has a higher mean experimental state fidelity, while the
mean waveform overlaps are similar on both processors.
The two shot counts were tested in separate sets of experiments on both processors. \texttt{ibm\_boston} used the same physical qubits at both shot counts, whereas \texttt{ibm\_pittsburgh} used different physical qubits between the two sets of experiments, so the differences do not isolate the effect of shot count for \texttt{ibm\_pittsburgh}.

\subsection{Scaling the preparation to a subsolar-mass waveform of $\mathcal{M}_c=0.2\,M_\odot$}

We now apply the same compact MPS preparation strategy to a subsolar-mass waveform with chirp mass $\mathcal{M}_c=0.2\,M_\odot$. This waveform contains
$36{,}782{,}082$ frequency nodes, approximately 143 times as many as
the GW170817 waveform. Multiband ROQ reduces this to 478 empirical
frequency nodes by using multiband PyROQ \citep{Qi2021_PyROQ, Newell2026_MultibandedROQ}.  This register requires a 9 qubits.

Exact preparation of the complex ROQ node vector requires 502 logical
CNOT gates. On \texttt{ibm\_boston}, the experimental state fidelity
estimated from the echo is reported as a $95\%$ upper limit of
$0.024$, while the reconstructed waveform overlap is $0.692$.
Applying MPS directly also requires a large circuit. An approximation
with maximum bond dimension 8 achieves a theoretical state fidelity
of only \(0.87\), yet requires 855 logical CNOT gates in our sequential compilation, compared with 502 for exact preparation. 

Combined amplitude and phase preconditioning enables an MPS approximation with maximum bond dimension 2. The sequential implementation requires 27 logical CNOT gates and one auxiliary bond qubit. On \texttt{ibm\_boston}, this circuit achieves an experimental state fidelity of $0.661$ and a reconstructed waveform overlap of $0.981$.
To further reduce the circuit cost, we construct the compact MPS preparation using 9 qubits and 13 logical CNOT gates, without an auxiliary qubit. This preparation achieves a theoretical state fidelity of $0.999994$. 

\begin{table}[!htbp]
\centering 
\caption{\textbf{Performance of six sets of $25$ experiments.} Values are
means $\pm$ standard deviations. The preparation uses $13$ CZ gates, with $37$ physical rotation
pulses for GW170817 and $39$ for the subsolar waveform. Echo estimates are unconditioned, and
node fidelities and waveform overlap condition on the valid node indices.
$F_{\rm echo}$ and $F_{\rm node}$ denote encoded and restored node-magnitude
fidelities. $O$ is  waveform accuracy. }
\label{tab:subsolar_results}
\footnotesize
\setlength{\tabcolsep}{14pt}
\begin{tabular}{llccccc}
\toprule
Processor & Shots & $F_{\mathrm{echo}}$ & $F_{\mathrm{node}}$
& $O$ & Discarded (\%)  \\
\midrule
\addlinespace[3pt]
\multicolumn{6}{c}{%
\textbf{GW170817}\quad
(8 qubits, 135 ROQ nodes, $\mathcal{M}_c=1.2\,M_\odot$)} \\
\addlinespace[5pt]
\texttt{ibm\_boston} & 16,384
& $0.9154\pm0.0037$
& $0.9978\pm0.0004$
& $0.9890\pm0.0081$
& $2.00\pm0.14$ \\
\texttt{ibm\_pittsburgh} & 16,384
& $0.9196\pm0.0068$
& $0.9978\pm0.0003$
& $0.9912\pm0.0072$
& $2.31\pm0.18$ \\
\midrule
\addlinespace[10pt]
\multicolumn{6}{c}{%
\textbf{Subsolar}\quad
(9 qubits, 478 ROQ nodes, $\mathcal{M}_c=0.2\,M_\odot$)} \\
\addlinespace[5pt]
\texttt{ibm\_boston} & 16,384
& $0.9061\pm0.0047$
& $0.9924\pm0.0008$
& $0.9883\pm0.0089$
& $0.75\pm0.10$ \\
\texttt{ibm\_pittsburgh} & 16,384
& $0.8893\pm0.0062$
& $0.9923\pm0.0008$
& $0.9862\pm0.0105$
& $0.90\pm0.07$ \\
\texttt{ibm\_boston} & 65,536
& $0.9108\pm0.0050$
& $0.9980\pm0.0002$
& $0.9963\pm0.0020$
& $0.74\pm0.05$ \\
\texttt{ibm\_pittsburgh} & 65,536
& $0.8879\pm0.0084$
& $0.9978\pm0.0002$
& $0.9973\pm0.0011$
& $1.11\pm0.08$ \\
\bottomrule
\end{tabular}
\end{table}

Our hardware compilation uses 13 physical CZ gates in 7 entangling
layers for the compact MPS. Like for GW170817, we reduce the number of physical rotation pulses without changing the prepared state. Relative to a direct
compilation of the compact circuit, the number of $\sqrt{X}$ pulses
decreases from 61 to 39, and the number of rotation pulse layers
decreases from 14 to 10.

Figure~\ref{fig:subsolar_amplitudes} shows the effect of amplitude
preconditioning on waveform reconstruction. Exact preparation and
phase preconditioning alone overestimate the small waveform magnitudes
at higher frequencies. Amplitude preconditioning improves their recovery,
as shown by the comparison of the two exact preparation circuits at the
same shot count. Combined amplitude and phase preconditioning with compact
MPS preparation reproduces the decrease in waveform magnitude across the
displayed frequency band using only 13 logical CNOT gates.

\begin{figure}[!htbp]
\centering
\includegraphics[width=\textwidth]{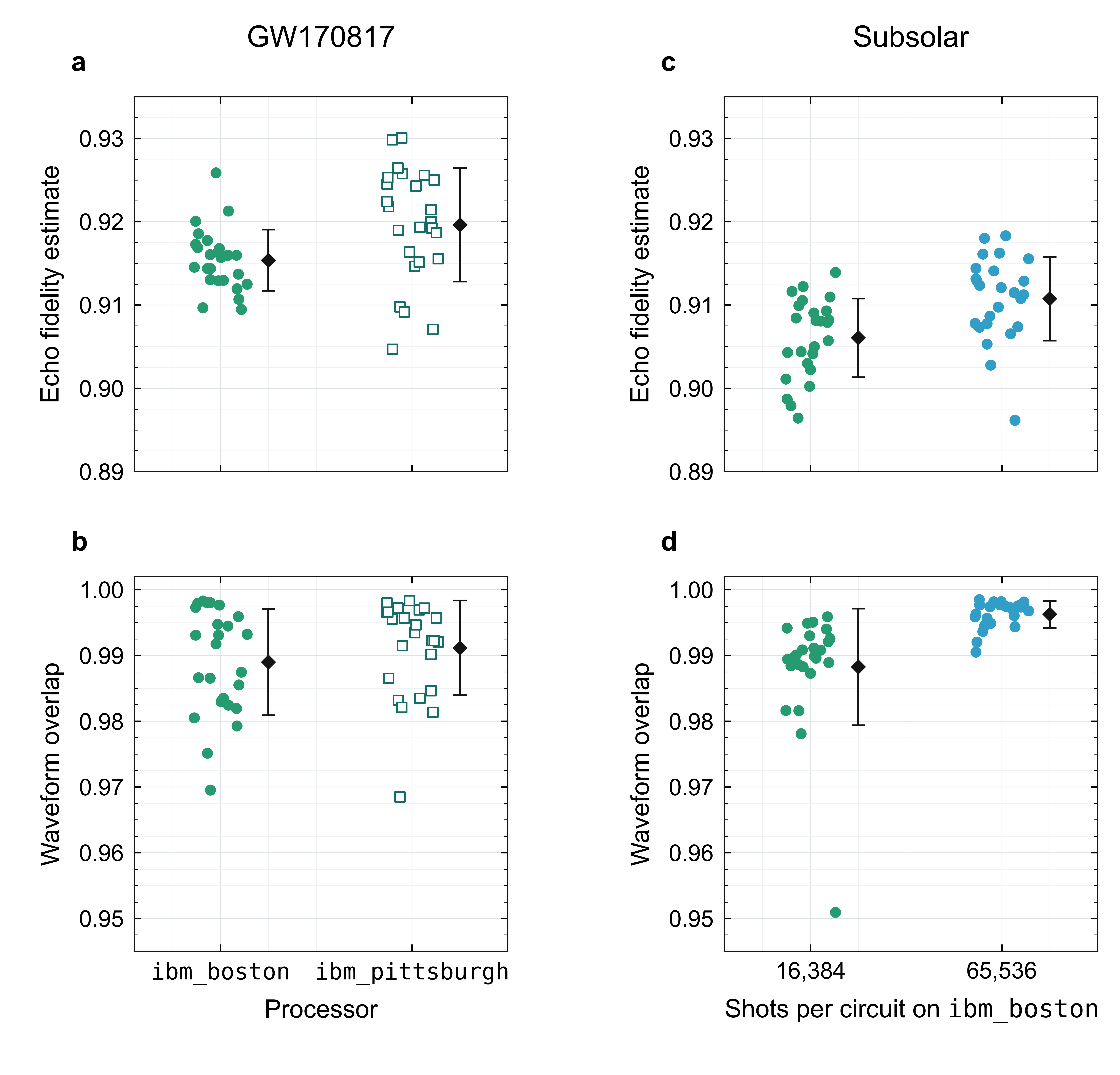}
\caption{\textbf{Repeated preparation of both waveforms on IBM Heron.}
Each point
is one of $25$ jobs at that setting. Filled circles denote \texttt{ibm\_boston} and open squares
\texttt{ibm\_pittsburgh}; small horizontal offsets distinguish coincident points. Black diamonds
and error bars show the mean and sample standard deviation. 
Increasing subsolar shots is
associated with higher overlap and less variation, while the echo estimates remain
well below the ideal state fidelities. 
}
\label{fig:campaigns}
\end{figure}

Figure~\ref{fig:campaigns} shows the individual fidelity and accuracy performances for 4 sets of 25 experiments. The 2 sets on the left for GW170817 show the results on \texttt{ibm\_boston} and \texttt{ibm\_pittsburgh}, respectively, using 16,384 shots. Both processors achieve similar mean state fidelities and waveform overlaps, although
\texttt{ibm\_boston} has less variation in state fidelity among experiments.  The 2 sets on the right for the subsolar waveform are on \texttt{ibm\_boston} with different shot counts.  The experiments with $65{,}536$ shots have higher reconstructed waveform accuracies and smaller variations than those with $16{,}384$ shots on both quantum processors, as expected. The experimental state fidelities with $16{,}384$ shots  are similar to those with $65{,}536$ shot counts. This feature is consistent with a contribution from finite sampling to waveform reconstruction error, while increasing the shot count does not change the preparation circuit.

\section{Discussion}\label{sec:discussion}
In this work, we have combined ROQ representation, invertible amplitude and phase preconditioning, and MPS compression to encode complete, long gravitational
waveforms as
quantum states using small registers on quantum hardware. ROQ drastically reduces the size of the waveform complex vector while preserving its accuracy, and preconditioning significantly
reduces the maximum bond dimension required for accurate MPS
preparation. Reducing local rotation pulses preserves the prepared states and enables compact MPS. In combination, these steps allow the GW170817 and subsolar
waveforms to be represented using
$135$ and $478$ empirical frequency nodes and encoded using 8 and 9 qubits, respectively, on IBM quantum processors with
13 physical CZ gates for each preparation. The subsolar result
demonstrates that a much longer waveform does not require a
proportionately more complex circuit, because the structure of the waveform
after ROQ compression and preconditioning determines the preparation cost.

Hardware execution and measurement are the main sources of
experimental error. The theoretical state fidelities exceed 0.9998,
whereas the experimental estimates from echoes are near 0.9.
Increasing the number of measurement shots reduces statistical errors in waveform reconstruction, but it does not improve the underlying quantum
state. It cannot remove gate errors or readout biases. Similarly, this conclusion holds for the number of echo shots. 
The $6$ sets of $25$ experiments on \texttt{ibm\_boston} and \texttt{ibm\_pittsburgh} provide an accuracy statistics of state preparation.

This state preparation method applies to arbitrary waveforms generated by LALSuite, a software package used by the gravitational wave community. The resource requirements depend on the waveform model, parameter range, and required accuracy. A next step is to use the preconditioned states
in quantum algorithms for matched filtering and Bayesian parameter
inference. Restoring the reference phase coherently would require a diagonal unitary. Alternatively, using the preconditioned states directly to formulate these subsequent algorithms could
avoid reconstructing every frequency node classically.

State preparation  does not provide a quantum computational advantage
\cite{Aaronson2015}. A useful comparison is the total
cost of a quantum analysis using the data encoding against an appropriate classical
method. Our present work provides an experimentally tested waveform state preparation procedure on which a subsequent quantum algorithm 
can build.

Beyond gravitational waves, this method may also be used to reduce data loading costs in other data-intensive research such as financial data. Quantum state preparation has been investigated for financial
time series \citep{Nakaji2022_FinancialEncoding,Jumade2023_DataLoading,Szoldra2026_MPSPreparation}
and for probability distributions used in pricing and risk
analysis \citep{Zoufal2019_DataLoading,Woerner2019_QuantumRisk}.  The current results motivate testing this combined method for other large datasets whose structure permits accurate compression.

\section{Methods}\label{sec:methods}

\subsection{Waveform models and reduced bases}

The perturbation of spacetime caused by gravitational waves produced by compact binary coalescences are recorded in the detectors as a strain $h$, which has two polarizations, plus ($h_+$) and cross ($h_\times$), and each is a complex vector. Gravitational waveform models describe the strain in terms of component masses, spins, and tidal deformabilities, together with parameters such as sky location. In this work, we use the
IMRPhenomPv2\_NRTidalv2 model \citep{Dietrich2019_NRTidalv2} in LALSuite \citep{lalsuite} to generate frequency-domain gravitational waveforms for their preparation in the quantum computers, and these complex vectors are normalized to unit
Euclidean norm when necessary. 

We use reduced bases and empirical interpolation to represent
gravitational waveforms accurately using a small set 
of frequencies named empirical frequency nodes. We refer to the reduced bases used
for this ROQ representation as ROQ bases. Each ROQ basis can
represent both polarizations, $h_+(f)$ and $h_\times(f)$, using the same
empirical frequency nodes \citep{Field2014_Surrogate,Qi2021_PyROQ}.

The first encoded waveform is GW170817, the first
binary neutron star merger detected through both gravitational waves and electromagnetic waves in 2017
\citep{Abbott2017_GW170817}. For GW170817, the reduced basis was constructed by singular value
decomposition of 200 normalized training waveforms with chirp masses
between $1.18$ and $1.22\,M_\odot$. The training set spans component
mass ratios $m_2/m_1\in[0.7,1]$ and tidal deformabilities
$\Lambda_{1,2}\in[0,1000]$, with aligned dimensionless spins fixed at
$\chi_1=\chi_2=0.05$. We kept singular vectors whose singular values exceeded
$10^{-8}$ times the largest singular value. The resulting basis
contains 135 elements. We then selected 135 empirical frequency nodes
by successively locating the largest interpolation residual of each
basis element. Subsequently, waveform reconstruction follows the empirical
interpolation method described in \cite{Field2014_Surrogate}. The waveform encoded for GW170817 has
$\mathcal{M}_c=1.1977\,M_\odot$, component mass ratio
$m_2/m_1=0.86$, aligned dimensionless spins
$\chi_1=\chi_2=0.05$, and tidal deformabilities
$\Lambda_1=\Lambda_2=425$. The theoretical mismatch of the ROQ representation with the original waveform is $1.5\times 10^{-17}$.

For the subsolar waveform, we used a reduced basis constructed with
multibanded PyROQ \citep{Qi2021_PyROQ,Newell2026_MultibandedROQ}.
PyROQ builds up a basis by selecting the training waveform with the largest empirical interpolation error and orthonormalizing it against
the existing basis until the basis can span the parameter range at targeted accuracy. The search proceeds through progressively larger
training subsets until the interpolation errors satisfy the targeted
tolerance. Multiband PyROQ applies this search method to multibanded waveforms adjusted through the frequency bands, rather than waveforms represented by frequencies uniformly distributed across the bands. Multibanding reduces the number of frequency points used
during this search by using finer frequency spacing at low frequencies
and coarser spacing at high frequencies to preserve the waveforms accurately. The subsolar basis was constructed over $20$--$9000$~Hz using the 
$4096$-s waveforms for chirp masses
$\mathcal{M}_c\in[0.1995,0.2005]\,M_\odot$, mass ratios
$m_1/m_2\in[1,5]$, and tidal deformabilities $\Lambda_{1,2}\in[0,5000]$. It has 478 linear empirical frequency nodes which range $20$--$5544$~Hz. The waveform encoded here has
$\mathcal{M}_c=0.2\,M_\odot$, equal component masses, aligned
dimensionless spins $\chi_1=\chi_2=0.05$, and tidal deformabilities
$\Lambda_1=\Lambda_2=500$. The theoretical mismatch of its ROQ representation with the original waveform is $1.2\times 10^{-7}$.

\subsection{Amplitude and phase preconditioning}

Suppose there are $N_L$ empirical frequency nodes for an ROQ basis. At each ROQ frequency $F_j$, the complex waveform value is
$h_j=h_+(F_j)=A_j e^{i\Psi_j}$, where $A_j$ and $\Psi_j$ are the
magnitude and phase obtained from LALSuite. 
Amplitude
preconditioning multiplies these values by $F_j^{7/6}$ to rescale
the leading inspiral amplitude dependence. Phase preconditioning
multiplies them by $e^{-i\Psi_{\mathrm{ref}}(F_j)}$ to remove the
reference chirp phase. The combined transformation is
\begin{equation}
\widetilde h_j
=
F_j^{7/6}e^{-i\Psi_{\mathrm{ref}}(F_j)}h_j.
\end{equation}
For the waveform at each frequency, we use its phase calculated by LALSuite as the reference phase, removing it before quantum state preparation.
The combined transformation therefore produces a real vector before
MPS approximation. The individual amplitude and phase preparations
use only the corresponding transformation.

These transformations are evaluated classically before circuit
construction. The transformed vector is normalized to unit
Euclidean norm,
\begin{equation}
\alpha_j
=
\frac{\widetilde h_j}
{\left(\sum_{k=1}^{N_{\mathrm L}}
|\widetilde h_k|^2\right)^{1/2}},
\end{equation}
and padded with zeros to length $2^n$, where
$n=\lceil\log_2N_{\mathrm L}\rceil$. Its amplitude encoding is
\begin{equation}
|\psi_\alpha\rangle
=
\sum_{j=1}^{N_{\mathrm L}}\alpha_j|j\rangle,
\end{equation}
where $|j\rangle$ is the computational basis state assigned to
the ROQ frequency node $F_j$.  The amplitude factors, reference phase, and normalization constants
are stored classically to reverse the transformation. The exponent
$7/6$ and the ordering of the empirical frequency nodes remain fixed
throughout the circuit comparisons and hardware experiments.

\subsection{Exact and MPS state preparations}

Exact preparation circuits are built for the normalized complex
vectors using Qiskit's state preparation routine and compiled
into logical CNOT gates and local rotations
\cite{Mottonen2005,Iten2016_Isometries}. The resulting circuits contain
247 logical CNOT gates for the 135-node GW170817 vector and 502 for
the 478-node subsolar vector.

For MPS preparation, the normalized vector is reshaped into a tensor
with one binary index per data qubit. Successive singular value
decompositions produce an MPS, retaining at most $\chi$ singular
values at each partition \cite{Oseledets2011}. The largest retained
dimension is the maximum bond dimension. The sequential construction
implements the resulting isometries using an auxiliary bond register
\cite{Schon2005,Ran2020}. For $\chi=2$, this register contains one
qubit, and the circuits require 24 and 27 logical CNOT gates for
GW170817 and the subsolar waveform, respectively. Theoretical state
fidelities are calculated from the circuit state vectors without
device noise, relative to the corresponding normalized vectors
before MPS truncation.

We also construct the MPS without a separate auxiliary register,
which we call compact MPS preparation. Each block is prepared from a
Schmidt decomposition across its central bond, followed by operations
on neighboring pairs of data qubits that extend the state towards
both ends. The GW170817 construction uses one block of eight qubits.
For the subsolar waveform, a further truncation to bond dimension
one separates an 8-qubit block from the remaining qubit, which
requires only a local rotation. We select the partition with the
lowest estimated CNOT count among contiguous partitions satisfying
a theoretical state fidelity of at least $0.999$ and a reconstructed
waveform overlap of at least $\sqrt{0.999}$. Both selected circuits
contain 13 logical CNOT gates, using eight and nine data qubits,
respectively.

For execution on IBM Heron processors, the compact circuits are
compiled into CZ gates, $\sqrt{X}$ pulses and virtual $Z$ rotations.
The circuit construction specifies each isometry only on the
states occupied during preparation, allowing its action on the
remaining states to be chosen during compilation
\cite{Iten2016_Isometries}. We use real rotation templates with at
most two CZ gates per isometry, remove redundant local pulses and
refit the remaining rotation angles to reproduce the same
approximate MPS. A compilation is accepted when the Euclidean
difference between the prepared state vectors, after alignment of
their global phases, is below $10^{-10}$. This optimization uses
theoretical state vectors, without a device noise model.

Both hardware preparations contain 13 physical CZ gates in seven
entangling layers. Physical rotation pulses are counted as
$\sqrt{X}$ operations; virtual $Z$ rotations are implemented through
changes to the control phase \cite{McKay2017_VirtualZ}. Rotation and
entangling layers are counted from the circuit dependencies
restricted to the corresponding operations. These counts precede
the additional operations introduced by runtime twirling and
dynamical decoupling.

\subsection{Hardware execution and analysis}

The hardware experiments use the IBM Heron processors
\texttt{ibm\_boston} and \texttt{ibm\_pittsburgh}. For each set of
repeated experiments, we select a connected chain of eight or nine
physical qubits by minimizing a calibration score,
\begin{equation}
S(\pi)
=
\sum_i r_{\pi(i)}
+
\sum_i n_i^{\mathrm{SX}}e_{\pi(i)}^{\mathrm{SX}}
+
\sum_{(i,j)} n_{ij}^{\mathrm{CZ}}
e_{\pi(i),\pi(j)}^{\mathrm{CZ}},
\end{equation}
where $\pi$ maps logical qubits to physical qubits, $r$ is the
readout error, and $e^{\mathrm{SX}}$ and $e^{\mathrm{CZ}}$ are the
calibrated errors of the $\sqrt{X}$ and CZ gates. The weights
$n_i^{\mathrm{SX}}$ and $n_{ij}^{\mathrm{CZ}}$ are their occurrence
counts in the preparation circuit. The calibration score is used for qubit selection; the experimental
fidelities are calculated from measured outcomes. The selected qubits and compiled
circuits remain fixed throughout each set of 25 experiments.

The comparisons of exact preparation and the individual and combined
preconditioning use $16{,}384$ shots per circuit on
\texttt{ibm\_boston}, with sampling and echo circuits submitted in
separate jobs. For the repeated compact MPS experiments, each job
contains one sampling circuit and one echo circuit. GW170817 uses $16{,}384$ shots per circuit on each processor,
giving 2 sets of 25 experiments. The subsolar waveform
uses sets with $16{,}384$ and $65{,}536$ shots per circuit
on each processor, giving 4 sets of 25 experiments. These repeated
experiments use gate and measurement twirling with 32 randomizations
and XX dynamical decoupling.

The echo circuit consists of the preparation circuit followed by
its inverse, with a barrier between them to prevent cancellation
during compilation. For both waveforms, the compact MPS echo circuit contains
26 physical CZ gates, comprising 13 in the preparation
circuit and 13 in its inverse. Let $P_0$ be the probability of obtaining the all-zero bit string
when measuring the echo circuit. We estimate the experimental state
fidelity as
\begin{equation}
F_{\mathrm{exp}}
=
\sqrt{
\max\left(
0,\frac{P_0-2^{-n}}{1-2^{-n}}
\right)},
\end{equation}
where $n$ is the total number of measured qubits, including any
auxiliary qubits. The correction removes the return probability
$2^{-n}$ of a fully mixed state, and the square root accounts for
the preparation and inverse circuits. Interpreting this quantity
as a preparation fidelity estimate assumes approximately
depolarizing noise and comparable errors in the two circuits.
All echo outcomes enter the estimate. Readout and initialization
errors are not separately corrected. We use the measured fraction $\widehat P_0$ and the total
number of echo shots in a binomial test to determine whether
$P_0$ exceeds $2^{-n}$ at the $3\sigma$ significance level.
If the excess is not significant, we calculate an exact
$95\%$ upper confidence bound on $P_0$ and substitute this
bound for $\widehat P_0$ in the fidelity formula to obtain
an upper limit on $F_{\mathrm{exp}}$.

The sampling circuit measures the prepared state in the computational
basis. Outcomes assigned to the zero padding are excluded from
waveform reconstruction. For sequential MPS preparation, we also
require the auxiliary bond qubit to have measurement outcome zero.
The discarded fraction is the number of excluded outcomes divided
by the total number of sampling shots. These exclusions apply only
to waveform reconstruction and the associated magnitude fidelity.

Let $C_j$ be the number of accepted sampling outcomes assigned to
the empirical frequency node $F_j$, with
$j=1,\ldots,N_{\mathrm L}$. The probabilities used to recover the
waveform magnitudes are
\begin{equation}
p_j
=
\frac{C_j}{\sum_{k=1}^{N_{\mathrm L}}C_k}.
\end{equation}
Writing $w_j=F_j^{7/6}$ when amplitude preconditioning is used and
$w_j=1$ otherwise, we reverse the amplitude rescaling and restore
the original norm,
\begin{equation}
\widehat A_j
=
\frac{
\|\boldsymbol h\|_2\,w_j^{-1}\sqrt{p_j}
}{
\left(
\sum_{k=1}^{N_{\mathrm L}}w_k^{-2}p_k
\right)^{1/2}
},
\end{equation}
where $h=(h_1,\ldots,h_{N_{\mathrm L}})$ contains the
theoretical waveform values at the empirical frequency
nodes.

The restored waveform magnitude fidelity at these nodes is
\begin{equation}
F_{\mathrm{node}}
=
\frac{
\left(
\sum_{j=1}^{N_{\mathrm L}}\widehat A_j|h_j|
\right)^2
}{
\left(
\sum_{j=1}^{N_{\mathrm L}}\widehat A_j^2
\right)
\left(
\sum_{j=1}^{N_{\mathrm L}}|h_j|^2
\right)
}.
\end{equation}
This quantity measures agreement between the recovered and
theoretical magnitudes and is insensitive to relative phase errors.

We reconstruct the waveform by combining the recovered magnitudes
with the known model phases and applying empirical interpolation,
\begin{equation}
h_{\mathrm{rec}}(f)
=
\sum_{j=1}^{N_{\mathrm L}}
B_j(f)\widehat A_j e^{i\Psi_j},
\end{equation}
where $B_j(f)$ are the interpolation functions associated with the
ROQ representation and $\Psi_j$ are the waveform phases defined
in Section~4.2. These phases are supplied by the waveform model,
rather than recovered from quantum measurements.

The normalized waveform overlap (also called waveform accuracy) is
\begin{equation}
O
=
\frac{
\left|
\sum_k h_{\mathrm{ref}}^*(f_k)h_{\mathrm{rec}}(f_k)
\right|
}{
\left[
\left(\sum_k|h_{\mathrm{ref}}(f_k)|^2\right)
\left(\sum_k|h_{\mathrm{rec}}(f_k)|^2\right)
\right]^{1/2}
},
\label{eq:overlap}
\end{equation}
where $h_{\mathrm{ref}}$ is the comparison waveform and $f_k$ are
the evaluation frequencies. The waveform mismatch is defined as $1-O$.

For GW170817, $h_{\mathrm{ref}}$ is the original theoretical waveform
on $257{,}025$ uniformly spaced frequency points from 20 to
1024~Hz. For the subsolar waveform, it is the classical ROQ
reconstruction of the theoretical node values on $266{,}060$
multiband frequency points extending from 20~Hz to around 5544.25~Hz, which has a mismatch of $1.23\times10^{-7}$ with its original waveform. Each evaluation point receives equal weight, without
weighting by a detector noise spectrum or by frequency spacing.

Each fidelity and overlap is calculated separately for each
experiment. Reported averages are the means of these individual
values, and uncertainties are standard deviations for
the 25 experiments.

\backmatter

\bmhead{Data availability}
The reduced bases, encoded circuits, and measured results are available from the
author on reasonable request.

\bmhead{Code availability}
Encoding and analysis code is available from the author on reasonable request.

\bibliographystyle{sn-nature}
\bibliography{main}

\vspace{10pt}
\noindent\textbf{Acknowledgements\\}
H.Q. thanks Frank Linde for inspiring her to pursue quantum computing for gravitational wave astronomy in 2021. This project was funded and supported by the UK National Quantum Computer Centre [NQCC200921], which is a UKRI Centre and part of the UK National Quantum Technologies Programme (NQTP).  This project is based on work previously built on the computational resources provided by LIGO Laboratory and the Leonard E Parker Center for Gravitation, Cosmology and Astrophysics at the University of Wisconsin-Milwaukee and supported by National Science Foundation Grants PHY-0757058, PHY-0823459, PHY-1626190, and PHY-1700765. 

\vspace{10pt}
\noindent\textbf{Author contributions\\}
H.Q. conceived and designed the study, developed the methods, conducted the experiments on IBM quantum computers, analyzed the data and interpreted the results, and wrote the manuscript.

\end{document}